\documentclass{jfm}
\usepackage{graphicx}
\usepackage{epstopdf, epsfig}

\usepackage{amsmath}
\usepackage{dcolumn}    
\usepackage{bm}         
\usepackage[IT]{subfigure}
\usepackage{siunitx}    
\usepackage{mathrsfs}       
\usepackage{color}
\let\Gamma\varGamma
\usepackage[normalem]{ulem}

\newcommand{\ka}{{k_a}}
\newcommand{\kd}{{k_d}}
\newcommand{\kp}{{k}}
\newcommand{\mi}{{m_{\mathrm{I}}}}
\newcommand{\bk}[2]{{b_{#1}^{(#2)}}}
\newcommand{\ck}[2]{{c_{#1}^{(#2)}}}
\newcommand{\ept}[2][]{\exp(-p_{#2}^2 t_{#1})}

\newcommand{\lbk}{\left(}
\newcommand{\rbk}{\right)}
\newcommand{\nmv}{\mathcal{V}}
\newcommand{\nmD}{\mathcal{D}}
\newcommand{\tr}{E}

\newcommand{\RNum}[1]{\uppercase\expandafter{\romannumeral #1\relax}}
\newcommand{\Pe}{\mbox{\textit{Pe}}}

\newcommand{\hl}[1]{\textcolor{black}{#1}}
\newcommand{\cor}[1]{\textcolor{black}{#1}}
\newcommand{\del}[1]{}

\shorttitle{Transient solute transport with sorption}
\shortauthor{L. Zhang, M. A. Hesse and M. Wang}

\title{Transient solute transport with sorption in Poiseuille flow}

\author{Li Zhang\aff{1,2},
  Marc A. Hesse\aff{2,3} \corresp{\email{mhesse@jsg.utexas.edu}}
 \and Moran Wang\aff{1}}

\affiliation{\aff{1}Department of Engineering Mechanics and CNMM, Tsinghua University, Beijing, 100084, China \aff{2} Department of Geological Sciences, University of Texas at Austin, Austin, TX, 78712, US \aff{3} Institute of Computational Engineering and Sciences, University of Texas at Austin, Austin, TX, 78712, US}

\begin{document}

\maketitle

\begin{abstract}
Previous work on solute transport with sorption in Poiseuille flow has reached contradictory conclusions. Some have concluded that sorption increases mean solute transport velocity and decreases dispersion relative to a tracer, while others have concluded the opposite. Here we resolve this contradiction by deriving a series solution for the transient evolution that recovers previous results in the appropriate limits. This solution shows a transition in solute transport behavior from early to late time that is captured by the first- and zeroth-order terms. Mean solute transport velocity is increased at early times and reduced at late times, while solute dispersion is initially reduced, but shows a complex dependence on the partition coefficient $\kp$ at late times. In the equilibrium sorption model, the time scale of the early regime and the duration of the transition to the late regime both increase with $\ln \kp$ for large $\kp$. The early regime is pronounced in strongly-sorbing \del{and advection-dominated} systems ($\kp\gg1$\del{, $\Pe \gg1$}). The kinetic sorption model shows a similar transition from the early to the late transport regime and recovers the equilibrium results when adsorption and desorption rates are large. As the reaction rates slow down, the duration of the early regime increases, but the changes in transport velocity and dispersion relative to a tracer diminish. In general, if the partition coefficient $\kp$ is large, \cor{the early regime is well-developed} and the behavior is well characterized by the analysis of the limiting case without desorption.
\end{abstract}


\section{Introduction}
    Reactive solute transport with surface reaction is common in natural and engineering applications such as solute separation in chromatography \citep{Hlushkou2014}, contaminant transport in porous media \citep{Hesse2010} and particle transport in biological systems \citep{Shipley2012}. Solute transport in a channel with Poiseuille flow and sorbing boundaries provides a simplified model system that allows an understanding of the effect of reactions on the macroscopic transport velocity and dispersion of the solute. This two-dimensional configuration resembles some microfluidic systems used in chromatography and biomaterial delivery and provides insight of solute transport in fractures \citep{Wels1997}. In these systems, the macroscopic properties are given by transverse averaging. In the absence of surface reactions, the solute is a tracer and the average transport velocity of the tracer is identical to the mean flow velocity and its dispersion is given by Taylor's analysis \citep{Taylor1953}. However, previous work has reached contradictory conclusions as to the effect of sorption on solute transport velocity and dispersion.
    
    \hl{ For channel flow with first-order, \emph{irreversible} adsorption reaction, previous analyses have shown that adsorption increases transport velocity and decreases the dispersion of the solute relative to a tracer in the asymptotic regime \citep{SanGill1973, deGance1978b, deGance1978a, Lungu1982, Smith1983, Barton1984, Shapiro1986, Balakotaiah1995, Mikelic2006, Biswas2007}}. The solute velocity increases because adsorption removes solutes from the slow-moving fluid near the wall so that the solute preferentially samples the fast-moving fluid in the center of the channel. This can increase the transport velocity by up to $30\%$ with increasing adsorption \hl{in planar Poiseuille flow} \citep{Lungu1982}. 
    
    However, this is in contrast to the results in chromatography showing that adsorption reduces the transport velocity due to the continuous removal of the solute from the concentration front \citep{Golay1958, Khan1962}. The chromatographic analysis considers a \emph{reversible} reaction that allows both adsorption and desorption. In this case, the transport of solute is determined by the partition coefficient $\kp$, the ratio of adsorbed mass over aqueous mass. Concretely, the transversely-averaged transport velocity will be reduced by a factor of $1/(1+\kp)$ relative to the mean flow velocity. Similarly, different results have been reached with respect to the effect of adsorption on the dispersion coefficient. \hl{Chromatographic analysis shows a complex dependence of dispersion on $\kp$ while dispersion is reduced in the former case.}
    
    The main difference between these two contrasting analyses is that one only considers adsorption \citep[e.g.][]{Lungu1982} while the other considers both adsorption and desorption \citep[e.g.][]{Khan1962}. One might therefore expect that the \cor{reversible analysis recovers the results of the irreversible one} in the limit of negligible desorption. However, in this limit the discrepancy between the two analyses is the largest. \cor{The transport velocity vanishes in the reversible case while it is finite in the irreversible case. This apparent contradiction may be reconciled by the observation that solute transport undergoes a transition from an early regime characterized by increased solute velocity to a late regime characterized by decreased solute velocity \citep{Paine1983, Balakotaiah1995}.}
    
    \cor{Here we present an analysis that demonstrates the transition in solute transport behavior reconciles the reversible and irreversible analyses.} To this end, we study solute transport in a two-dimensional straight channel with adsorption onto and desorption from the walls. We use the method of moments in combination with the Laplace transform to derive a set of series solutions for zeroth-, first- and second-order longitudinal moments valid for all times. It is shown that the zeroth-order terms in the series solution corresponds to the late time behavior, while the first-order terms corresponds to the early time behavior. This analysis recovers both the previous results and therefore reconciles them. Moreover, it allows us to quantify the transition for equilibrium and kinetic sorption models. The manuscript is structured as follows: the problem is formulated in \textsection 2 and solved in \textsection 3, followed by a discussion of the transport regimes in \textsection 4.

\section{Problem formulation}
        \begin{figure}
          \centerline{\includegraphics{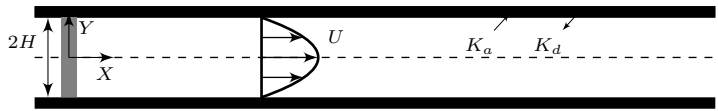}}
          \caption{In an infinitely long channel, a slender, transversely uniform strip of solute (gray area) is released in the fluid and transported by Poiseuille flow with adsorption and desorption on the walls.}
        \label{fig:phy_model}
        \end{figure}
    We model single component solute transport in a \hl{two-dimensional} straight channel with surface adsorption and desorption, which is illustrated in figure \ref{fig:phy_model}. The width of the channel is $2H$ in the $Y$ direction and the length is assumed to be infinite in the $X$ direction. The velocity field is given by an ideal Poiseuille flow, $U(Y) = \frac{3}{2} U_0 \left( 1 - Y^2/H^2 \right)$, where $U_0$ is the mean flow velocity. Adsorption onto and desorption from the walls allow exchange of mass between the solid surface and the fluid.
    
    The mass transport of solute in the fluid is given by the advection--diffusion equation,
        \begin{equation}
             \frac{\partial C}{\partial T} + U(Y) \frac{\partial C}{\partial X}  = D \left( \frac{\partial^2 C}{\partial X^2} + \frac{\partial^2 C}{\partial Y^2} \right),         \label{eq:concentration}
        \end{equation}
    where $T$ is the dimensional time [\SI{}{T}], $C$ the dimensional solute concentration [\SI{}{M L^{-3}}] and $D$ the diffusion coefficient [\SI{}{L^2 T^{-1}}].

    Since the channel is assumed to be infinite, the concentration and any order of its derivative must vanish as $X \rightarrow \pm \infty$. Because of the symmetry along the centerline, $\partial C/ \partial Y = 0$, and only the upper half of the domain is considered.
    
    The adsorbed concentration on the wall is assumed to form an infinitely-thin and static surface layer without longitudinal diffusion. The exchange of mass between the wall and the fluid is given as
        \begin{equation}
            -D \frac{\partial C}{\partial n} = \frac{\partial \Gamma}{\partial T} ,
        \end{equation}
    where $n$ denotes the outward normal direction of the wall and $\Gamma$ is the dimensional surface concentration [\SI{}{M L^{-2}}]. Adsorption and desorption are assumed to be described by the first-order reactions, so that the change of surface concentration is given by
        \begin{equation}\label{eq:kinetic_boundary}
            \frac{\partial \Gamma}{\partial T} = K_{a}C - K_{d}\Gamma,
        \end{equation}
    where $K_a$ and $K_d$ are the dimensional adsorption and desorption rate constants with dimensions of [\SI{}{L T^{-1}}] and [\SI{}{T^{-1}}], respectively \hl{\citep{Khan1962}}. When $K_d = 0$, \hl{the linear kinetic model reduces to a first-order, irreversible adsorption reaction}. When both $K_a$ and $K_d$ are large, the reaction approaches local chemical equilibrium. At equilibrium, the surface concentration is linearly proportional to solute concentration
        \begin{equation} \label{eq:lin_isotherm}
            \Gamma = K C ,
        \end{equation}
    where $K = K_a/K_d$ is the dimensional partition coefficient. Equation (\ref{eq:lin_isotherm}) is also referred to as a linear isotherm \citep{Golay1958}.

    Initially, the solute has a uniform transverse distribution at $X = 0$ with no mass adsorbed on the wall and is assumed to be a $\delta$-function in the $X$ direction so that
        \begin{subeqnarray}
            \gdef\thesubequation{\theequation \textit{a,b}}        
            C(X, Y, 0) = \frac{M_I}{A} \, \delta (X)  \quad \mathrm{and} \quad \Gamma(X, 0) = 0,
        \end{subeqnarray}
    where $M_I$ represents the total mass in the system [\SI{}{M}] and $A$ is the cross-sectional area of the channel [\SI{}{L^2}]. A characteristic concentration is chosen as $C_0 = M_I/(HA)$ to simplify the formulation in dimensionless form. This initial condition, which has been used in previous work, assumes that the system is not in local chemical equilibrium. We note that the transient solute transport behavior is very sensitive to the initial condition and further analysis of the effect of the initial condition is provided in appendix \ref{subsec:init_cd}.
    
    The following characteristic quantities are chosen to non-dimensionalize the problem,
        \begin{equation}
            \begin{array}{ccc}
                x = X/H, & y = Y/H, & u = U/U_0, \\
                c = C/C_0, & \gamma = \Gamma/(C_0 H), & t = T/(H^2/D).
            \end{array}
        \end{equation}
    Note that we choose $C_0 H$ as characteristic surface concentration and the diffusive time scale $H^2/D$ as the characteristic time scale. Consequently, the dimensionless formulation of the problem is written as 
        \begin{subequations} \label{eq:dml_eqs}
            \begin{align}
                \frac{\partial c}{\partial t} + \Pe \, u \frac{\partial c}{\partial x} & = \frac{\partial^2 c}{\partial x^2} + \frac{\partial^2 c}{\partial y^2}    \label{eq:dml_ade}\\
                -\frac{\partial c}{\partial y} & = \frac{\partial \gamma}{\partial t} \quad \mbox{at\ } \quad y = 1,   \label{eq:dml_bc_wall} \\
                \frac{\partial c}{\partial y} & = 0 \quad \mbox{at\ } \quad y = 0,   \label{eq:dml_bc_sym} \\
                c & = \delta (x) \quad \mbox{at\ } \quad t = 0, \label{eq:dml_ini} \\
                \cor{\gamma} & \cor{= 0 \quad \mbox{at\ } \quad t = 0.}
            \end{align}
        \end{subequations}
    If the surface reaction is modelled by the linear kinetic model, 
        \begin{equation} \label{eq:dml_knt_eq}
            \frac{\partial \gamma}{\partial t} = \ka c - \kd \gamma \,,
        \end{equation}
    there will be three dimensionless groups in equation (\ref{eq:dml_eqs}) and (\ref{eq:dml_knt_eq}),
        \begin{equation}
            \Pe = \frac{U_0 H}{D}, \quad \ka = \frac{K_a H}{D}, \quad \kd = \frac{K_d H^2}{D}.
        \end{equation}
    Physically, the Peclet number, $\Pe$, represents the ratio of the transverse diffusive time scale to the longitudinal advective time scale and the Damk\"{o}hler numbers, $k_a$ and $k_d$, represent the ratio of the transverse diffusive time scale to the adsorption and desorption time scales, respectively.
    
    Otherwise, if the equilibrium sorption model is used
        \begin{equation}
            \gamma = \kp c,
        \end{equation}
    and the number of dimensionless groups reduces to two by replacing $\ka$ and $\kd$ with
        \begin{equation}
            \kp = \frac{\ka}{\kd}.
        \end{equation}
    In this work, $\ka$, $\kd$ and $\kp$ are all assumed to be constants. \cor{A value of $10$ is chosen for $\Pe$ to limit the longitudinal domain size required in numerical simulation. This choice will not affect the key results which are independent of $\Pe$}. In the following, we deal with the more general linear kinetic sorption model analytically and results will be given for both the kinetic and equilibrium model in \textsection \ref{sec:results}.

\section{Solution for the longitudinal moments}
    \hl{Following the classical, transverse-averaging idea introduced by \citet{Taylor1953} to reduce the dimension of the problem, we consider the transverse-averaged concentration} $\bar{c} = \int_{0}^1 c\, dy $
    and the distribution of $\bar{c}$ is described by its longitudinal moments $m_n = \int_{-\infty}^{\infty} x^n \bar{c} dx$,
    where $n$ is the order of the moment. The lower-order moments, e.g. zeroth-, first- and second-order moments are of most interest to us. Furthermore, we define the normalized longitudinal moments of zeroth-, first- and second-order
        \begin{subeqnarray}
            \label{eq:M_def}
            \gdef\thesubequation{\theequation \textit{a,b,c}}
                M_0  = \frac{m_0}{\mi}, \quad
                M_1  = \frac{m_1}{m_0}, \quad
                M_2  = \frac{m_2}{m_0} - \lbk \frac{m_1}{m_0} \rbk^2,
        \end{subeqnarray}
    where $\mi$ is the dimensionless initial mass, which is unity here. The fraction of solute in the fluid is given by $M_0$. The center of mass and the variance of the solute distribution in the fluid are given by $M_1$ and $M_2$, respectively. Thus, the dimensionless transport velocity and longitudinal dispersion coefficient are
        \begin{subeqnarray}
            \label{eq:v_D_def}
            \gdef\thesubequation{\theequation \textit{a,b}}   v  = \frac{\mathrm{d} M_1}{\mathrm{d} t} \quad \mathrm{and} \quad
                D_L = \frac{1}{2} \frac{\mathrm{d} M_2}{\mathrm{d} t}.
        \end{subeqnarray}
    In the following, analytical solutions are derived for lower order moments $m_n (n = 0, 1, 2)$ in the form of series solutions.

    \subsection{Moment equation and solution in the Laplace space} \label{subsec:me_lt}
        Firstly, following the method of moments developed by \citet{Aris1956}, multiply equation (\ref{eq:dml_ade}) by $x^n$ and integrate in the $x$ direction to obtain the equation for $c_n^* (y, t)$,
            \begin{equation}
                \frac{\partial c_n^*}{\partial t} + \Pe \, u \int\limits_{-\infty}^\infty x^n \frac{\partial c}{\partial x} dx = \int\limits_{-\infty}^\infty x^n \frac{\partial^2 c}{\partial x^2} dx + \frac{\partial^2 c_n^*}{\partial y^2},
            \end{equation}
        where $c_n^* = \int_{-\infty}^{\infty} x^n c \, dx$ is the $n^{\text{th}}$ longitudinal moment of concentration in the filament through $y$, which is not yet transversely averaged. The moments $m_n$ introduced above are the transverse averages of $c_n^*$. After integration by parts and noting that \cor{the concentration and all of its derivatives} vanish at infinity, we have
            \begin{subequations} \label{eq:cn_star}
                \begin{align}
                \frac{\partial c_n^*}{\partial t} - n \Pe \, u \, c_{n-1}^* = n(n-1)c_{n-2}^* + \frac{\partial^2 c_n^*}{\partial y^2},
                \label{eq:moment_eq}
                \end{align}
        where $ c_{-1}^{*} = c_{-2}^{*} = 0$. Similarly, the boundary conditions (\ref{eq:dml_bc_wall}) and (\ref{eq:dml_bc_sym}) give
                \begin{align}
                    -\frac{\partial c_n^*}{\partial y} & = \frac{\partial \gamma_n^*}{\partial t} = \ka c_n^* - \kd \gamma_n^* \quad \mbox{at\ } \quad y =  1, \\
                    \frac{\partial c_n^*}{\partial y} & = 0 \quad \mbox{at\ } \quad y = 0,
                \end{align}
            \end{subequations}
        where $\gamma_n^*$ is defined as the $n^{\text{th}}$ longitudinal moment of the surface concentration.      

        Laplace transform in time reduces (\ref{eq:cn_star}) to a system of ordinary differential equations only involving the transformed variable
            \begin{equation}
                \skew3\hat{c}_n^*(y, s) = \mathscr{L}\{c_n^*\}(s) = \int\limits_0^\infty c_n^* e^{-st} dt,
            \end{equation}        
        because the transformed longitudinal moments of surface concentration $\skew3\hat{\gamma}_n^* = \mathscr{L}\{\gamma_n^*\}$ in the boundary condition can be eliminated. In Laplace space, equations (\ref{eq:cn_star}) are given by
            \begin{subequations} 
                \begin{align}
                \frac{\partial^2\skew3\hat{c}_n^*}{\partial y^2} & = s \skew3\hat{c}_n^* - c_n^* (t = 0)- n \Pe \, u \, \skew3\hat{c}_{n-1}^* - n(n-1)\skew3\hat{c}_{n-2}^*, \label{eq:Lp_moment_eq} \\
                -\frac{\partial \skew3\hat{c}_n^*}{\partial y} & = s\skew3\hat{\gamma}_n^* - \gamma_n^*(t = 0) =\ka \skew3\hat{c}_n^* - \kd \skew3\hat{\gamma}_n^* \quad \mbox{at\ } \quad y = 1, \label{eq:Lp_bc_wall}\\
                -\frac{\partial \skew3\hat{c}_n^*}{\partial y} & = 0 \quad \mbox{at\ } \quad y = 0. \label{eq:Lp_bc_sym}
                \end{align}
            \end{subequations}
        Since no mass is adsorbed on the wall initially, $\gamma_n^*(t=0) = 0$. A discussion of the more general initial conditions is given in appendix \ref{subsec:init_cd}. Note that the second equality in (\ref{eq:Lp_bc_wall}) can be solved for $\skew3\hat{\gamma}_n^*$ as
            \begin{equation} 
                \skew3\hat{\gamma}_n^* = \frac{\ka}{\kd + s} \skew3\hat{c}_n^*.
            \end{equation}
        so that (\ref{eq:Lp_bc_wall}) turns into a Robin-type boundary condition
            \begin{equation} \label{eq:Lp_bc_robin}
                -\frac{\partial \skew3\hat{c}_n^*}{\partial y} = \frac{\ka s}{\kd + s} \skew3\hat{c}_n^* \quad \mbox{at\ } \quad y = 1.
            \end{equation}
        The $\delta$-function initial distribution of solute leads to the following initial conditions
            \begin{align}
                c_0^*  = 1, \quad c_1^* = c_2^* = 0 \quad \mbox{at\ } \quad t = 0.
            \end{align}
Therefore, equation (\ref{eq:Lp_moment_eq}), together with boundary conditions (\ref{eq:Lp_bc_sym}) and (\ref{eq:Lp_bc_robin}), gives the following system of ordinary differential equations (ODEs) for $\skew3\hat{c}_0^*$, $\skew3\hat{c}_1^*$ and $\skew3\hat{c}_2^*$,
        \begin{subequations} \label{eq:ode}
            \begin{align}
                \frac{d^2 \skew3\hat{c}_0^*}{d y^2} & = s \skew3\hat{c}_0^* - 1, \\
                \frac{d^2 \skew3\hat{c}_1^*}{d y^2} & = s \skew3\hat{c}_1^* - \Pe \, u \, \skew3\hat{c}_0^*, \\
                \frac{d^2 \skew3\hat{c}_2^*}{d y^2} & = s \skew3\hat{c}_2^* - 2 \Pe \, u \, \skew3\hat{c}_1^* - 2\skew3\hat{c}_0^*,              
            \end{align}
        \end{subequations}
with boundary conditions
        \begin{subequations}
            \begin{align}
                 \frac{d \skew3\hat{c}_n^*}{dy} & = -\frac{\ka s}{\kd + s}\skew3\hat{c}_n^* \quad \mbox{at\ } \quad y = 1, \\
                \frac{d \skew3\hat{c}_n^*}{dy} & = 0 \quad \mbox{at\ } \quad y = 0,
            \end{align}
        \end{subequations}
for $n = 0, 1, 2$.

        
        However, not the analytical solutions of $\skew3\hat{c}_n^* (y, s)$, but the transverse-averaged moments $\skew3\hat{m}_n (s) = \int_0^1 \skew3\hat{c_n^*} dy$, are of interest here.
        For instance, $\skew3\hat{m}_0$ has the form
            \begin{equation}
                \skew3\hat{m}_0 = \frac{1}{s} - \frac{\ka \sinh \lbk \sqrt{s} \rbk}{\sqrt{s} \lbk \ka s \cosh \lbk\sqrt{s}\rbk+ \sqrt{s} \sinh\lbk \sqrt{s}\rbk \lbk \kd + s \rbk\rbk}.
            \end{equation}
        \cor{The analytical form of $\skew3\hat{m}_1$ and $\skew3\hat{m}_2$ are complex (given in supplementary materials)}, but both of them and $\skew3\hat{m}_0$ can be written in a general form as
            \begin{equation} \label{eq:moment_gnr}
                    \skew3\hat{m}_n(s) = \frac{N_n (s) }{\tr(s)^{(n+1)}} \quad \text{for} \quad n = 0, 1, 2,
            \end{equation}
        where the denominator $\tr(s)$ is given by
            \begin{equation}
                \tr(s) = \ka s \cosh \lbk {\sqrt{s}} \rbk + \left( \kd + s\right) \sqrt{s} \sinh \lbk {\sqrt{s}} \rbk ,
            \end{equation}
        which is a transcendental function of $s$ and includes all the singularities of the moments.
        The numerators $N_n(s)$ are complex functions of $s$ obtained by a computer algebra system \citep{mupad}. The transcendental function $\tr(s)$ has two important properties: 
        \begin{enumerate}
            \item There are only first-order singularities in $\tr(s)$, and thus $\skew3\hat{m}_0$, $\skew3\hat{m}_1$ and $\skew3\hat{m}_2$ only have first-, second- and third-order singularities, respectively. This helps to employ the residue theorem for the inverse Laplace transform.
            \item All the singularities of $\tr(s) = 0$ fall along the negative axis, and thus substituting $s = - p^2$, where $p$ is a real positive number, leads to a transcendental equation of $p$ in the real space \footnote{\cor{$\tanh(\mathrm{i} p) = \mathrm{i} \tan(p)$ is used.}},
            \begin{equation} \label{eq:tr_p_knt}
                \tan(p) (p^2 - \kd)- \ka p = 0.
            \end{equation}
            Equation (\ref{eq:tr_p_knt}) has an infinite number of roots $p_k, k = 0, 1, \dots, \infty$. These roots $p_k$ correspond to characteristic decay rates of the moments and the lowest order term with the smallest root, i.e. $p_0 = 0$, dominates the behavior at late times.
        \end{enumerate}

    \subsection{Inverse Laplace transform by the residue theorem} \label{subsec:ilt}
        The inverse Laplace transform of the moments can be written as the Bromwich integral
            \begin{equation}
                m_n(t) = \frac{1}{2 \upi \mathrm{i}} \int_{\mathcal{C}} \skew3\hat{m}_n(s) e^{st} ds,
            \end{equation}
        where $\mathrm{i} = \sqrt{-1}$ and $\mathcal{C}$ is a contour chosen so that all the singularities of $\skew3\hat{m}_n(s)$ are to the left of it. Further, if we apply the residue theorem to the above integral, we have
            \begin{equation} \label{eq:mn_Rk}
                m_n(t) = \sum\limits_{k=0}^{\infty} R_k,
            \end{equation}
        where $R_k$ are the residues of $\skew3\hat{m}_n e^{st}$ and can be calculated as
            \begin{equation} \label{eq:Rk_def}
                R_k = \frac{1}{(l-1)!} \lim_{s\to s_k} \frac{d^{l-1}}{ds^{l-1}} \lbk \skew3\hat{m}_n e^{st} (s-s_k)^{l} \rbk.
            \end{equation}
        where $l$ is the order of the $k^{\text{th}}$ singularity or pole $s_k$.

        Since $\skew3\hat{m}_0$, $\skew3\hat{m}_1$ and $\skew3\hat{m}_2$ only have first-, second- and third-order singularities respectively, we have
            \begin{subequations} \label{eq:moment_series}
                \begin{align} 
                    m_0(t) & = \sum\limits_{k=0}^{\infty} \lim_{s\to s_k} (s-s_k) \skew3\hat{m}_0 \exp(st) = \sum\limits_{k=0}^{\infty} a_k \exp(-p_k^2 t), \\
                    m_1(t) & = \sum\limits_{k=0}^{\infty} \lim_{s \to s_k} \frac{d}{ds} \left[ (s-s_k)^2 \skew3\hat{m}_1 \exp(st)\right] = \sum\limits_{k=0}^{\infty} b_k^{(1)} \exp(-p_k^2 t) + b_k^{(2)} t \exp(-p_k^2 t), \\
                    m_2(t) & = \sum\limits_{k=0}^{\infty} \frac{1}{2} \lim_{s \to s_k} \frac{d^2}{ds^2} \left[ (s-s_k)^3 \skew3\hat{m}_2 \exp(st)\right]  \nonumber \\ 
                           & = \sum\limits_{k=0}^{\infty} c_k^{(1)} \exp(-p_k^2 t) + c_k^{(2)} t \exp(-p_k^2 t) + c_k^{(3)} t^2 \exp(-p_k^2 t),
                \end{align}
            \end{subequations}
        where
            \begin{subequations} \label{eq:coeff_lim}
                \begin{align} 
                    a_k & = \lim_{s \to s_k} (s-s_k) \skew3\hat{m}_0, \\
                    b_k^{(1)} & = \lim_{s \to s_k} \frac{d}{ds}\left[ (s-s_k)^2 \skew3\hat{m}_1 \right], \\
                    b_k^{(2)} & = \lim_{s \to s_k} (s-s_k)^2 \skew3\hat{m}_1, \\
                    c_k^{(1)} & = \frac{1}{2} \lim_{s \to s_k} \frac{d^2}{ds^2} \left[ (s-s_k)^3 \skew3\hat{m}_2 \right], \\
                    c_k^{(2)} & = \lim_{s \to s_k} \frac{d}{ds} \left[ (s-s_k)^3 \skew3\hat{m}_2 \right],\\
                    c_k^{(3)} & = \frac{1}{2} \lim_{s \to s_k} (s-s_k)^3 \skew3\hat{m}_2.
                \end{align}
            \end{subequations}
        In order to remove the limit operator and give an explicit form of the coefficients in (\ref{eq:coeff_lim}), the general form of moments in Laplace space (\ref{eq:moment_gnr}) are substituted into (\ref{eq:coeff_lim}). The fractional forms of $\skew3\hat{m}_0$, $\skew3\hat{m}_1$ and $\skew3\hat{m}_2$ allow us to apply the L'Hospital's rule and obtain the explicit form of the coefficients,
            \begin{subequations} \label{eq:coeff_exp}
                \begin{align}
                    a_k & = \frac{N_0}{T_1}, \\
                    b_k^{(1)} & = \frac{T_1 N_1' - 2T_2N_1}{T_1^3}, \\
                    b_k^{(2)} & = \frac{N_1}{T_1^2}, \\
                    c_k^{(1)} & = \frac{(12 T_2^2 - 6 T_1 T_3) N_2- 6T_1 T_2 N_2' + T_1^2 N_2'' }{2T_1^5}, \\
                    c_k^{(2)} & = \frac{T_1 N_2' - 3T_2 N_2}{T_1^4}, \\
                    c_k^{(3)} & = \frac{N_2}{2T_1^3},
                \end{align}
            \end{subequations}
        where  $N_n' = dN_n/ds$, $N_n'' = d^2N_n/ds^2$ at $s = s_k$ and $ T_n = \tr^{(n)}/n !$ is the $n^{\text{th}}$ order Taylor expansion coefficient of $\tr(s)$ at $ s = s_k$. These coefficients can also be expressed in terms of $p_k$ by substituting $s_k = -p_k^2$. \cor{The analytical expressions of $a_k$, $b_k^{(1)}$, $b_k^{(2)}$, $c_k^{(1)}$,  $c_k^{(2)}$,  $c_k^{(3)}$ are given in the supplementary materials.}
        
        To summarize, for a given $\ka$ and $\kd$, equation \eqref{eq:tr_p_knt} is first solved for a series of $p_k$, which are substituted into \eqref{eq:coeff_exp} to obtain the coefficients $a_k$, $b_k$ and $c_k$. The normalized longitudinal moments $M_0$, $M_1$ and $M_2$, the transport velocity $v$ and the dispersion coefficient $D_L$ are then determined by definitions \eqref{eq:M_def} and \eqref{eq:v_D_def}.

    \subsection{Reduction to previous results} \label{sec:degeneration}
        In the long time limit, when the zeroth-order terms dominate, the transport velocity and dispersion coefficient are
            \begin{subequations} \label{eq:late_v_D}
                \begin{align} 
                    v_0 & = \frac{b_0^{(2)}}{a_0} = \Pe \frac{\kd}{\ka + \kd} = \Pe \frac{1}{\kp + 1}, \\
                    D_0 & = \frac{1}{2} \left( \frac{c_0^{(2)}}{a_0} - \frac{2b_0^{(1)} b_0^{(2)}}{{a_0}^2} \right) = \frac{1}{1 + \kp} + {\Pe}^2 \frac{2}{105} \frac{1 + 9\kp+ 25.5\kp^2}{(1 + \kp)^3} + \frac{\Pe^2}{\kd} \frac{\kp}{(1 + \kp)^3}, \label{eq:late_D}
                \end{align}
            \end{subequations}
        which are consistent with the results obtained in chromatography \citep{Khan1962}. For $k > 0$, the transport velocity of the solute is slower than the mean flow velocity at late times.

        At early but finite time, the first-order terms dominate and lead to an asymptotic velocity $v_1$ and dispersion coefficient $D_1$ given as
            \begin{subeqnarray} \label{eq:early_v_D}
            \gdef\thesubequation{\theequation \textit{a,b}}
                    v_1  =  \frac{\bk12}{a_1} \quad \mathrm{and} \quad D_1  = \frac{1}{2} \left( \frac{c_1^{(2)}}{a_1} - \frac{2b_1^{(1)} b_1^{(2)}}{{a_1}^2} \right).
            \end{subeqnarray}
        In the limiting case of $\kd = 0$ analysed by \citet{Lungu1982}, the zeroth-order coefficients of the moments vanish, i.e. $a_0 = b_0^{(2)} =  b_0^{(1)} = c_0^{(3)} = c_0^{(2)} = c_0^{(1)} = 0 $. Therefore, the first-order terms dominate and  lead to the following asymptotic transport velocity and dispersion coefficient,\footnote{There is a typo in $D_{\text{LM}}$ in published version. Here it has been corrected.}
            \begin{subequations} \label{eq:LM_v_D}
                \cor{\begin{align}
                    v_{\text{LM}} & = \frac{\Pe \left(4\, {{\ka}}^2\, {p_1}^2 + 3\, {{\ka}}^2 + 3\, {\ka} + 4\, p_1^4 - 3\, p_1^2\right)}{4\, p_1^2\, \left({{\ka}}^2 + {\ka} + p_1^2\right)}, \label{eq:LM_v} \\
                    D_{\text{LM}} & = 1 + \Bigl( \Pe^2 \bigl( -8\, \ka^6\, p_1^{4} + 150\, \ka^6\, p_1^2 - 315\, \ka^6 - 56\, \ka^5\, p_1^{4} + 750\, \ka^5\, p_1^2 \nonumber \\ 
                    &  - 945\, \ka^5 - 24\, \ka^4\, p_1^{6} + 282\, \ka^4\, p_1^{4} + 555\, \ka^4\, p_1^2 -945\, \ka^4 -192\, \ka^3\, p_1^{6} \nonumber \\
                    & + 1560\, \ka^3\, p_1^{4} -540\, \ka^3\, p_1^2 -315\, \ka^3 - 24\, \ka^2\, p_1^{8} -78\, \ka^2\, p_1^{6} + 1455\, \ka^2\, p_1^{4} \nonumber \\
                    & -495\, \ka^2\, p_1^2 -136\, \ka\, p_1^{8} + 810\, \ka\, p_1^{6} +225\, \ka p_1^{4} - 8\, p_1^{10} - 210\, p_1^{8} + 585\, p_1^{6} \bigr) \Bigr) \nonumber \\
                    & \Big/ \Bigl( 160\, p_1^{6} \bigl( \ka^2 + \ka + p_1^2 \bigr)^3 \Bigr), \label{eq:LM_D}
                \end{align}}
            \end{subequations}
        where $p_1$ is determined by solving (\ref{eq:tr_p_knt}). Equations (\ref{eq:LM_v_D}) are consistent with (3.4) and (3.11) given in \citet{Lungu1982}, except for a difference in notation. For $\ka > 0$, the asymptotic transport velocity of the solute is faster than the mean flow velocity. \cor{Note that the early and late transport velocity $v_1$, $v_0$ have linear dependence on $\Pe$ and the early and late dispersion coefficient $D_1$, $D_0$ (excluding contribution from pure diffusion) have quadratic dependence on $\Pe$ so that the normalized ones defined in \eqref{eq:nmD_def} below are generally independent of $\Pe$.}

    \subsection{Equilibrium sorption model} \label{sec:eq_sol}
        \cor{If the kinetics of the reactions are fast enough that local chemical equilibrium is valid, the linear kinetic sorption model reduces to the linear isotherm \cor{(i.e., equilibrium sorption model)} $\gamma = \kp c$, with $\kp = \ka/\kd$. For the equilibrium sorption model, equation (\ref{eq:tr_p_knt}) becomes
            \begin{equation} \label{eq:tr_p_eq}
                \tan(p) = - \kp \, p,
            \end{equation}
        which can be solved for a series of $p_k$. Taking the limit $\ka \rightarrow \infty, \kd \rightarrow \infty$ of (\ref{eq:coeff_exp}) while keeping $\ka/\kd = \kp$, the coefficients become only functions of the partition coefficient $\kp$, as expected.}

    \subsection{First-order approximation of the series solution}
        For the general case when $\kd$ is not zero, the fast transport described by (\ref{eq:LM_v_D}) may survive at early times before desorption has come into play. In this case, the general series solution of the moments (\ref{eq:moment_series}) allows us to study the transition from fast transport at early times described by first-order terms to slow transport at late times described by zeroth-order terms.
        
\begin{figure}
            \centering
            \subfigure{
                \includegraphics{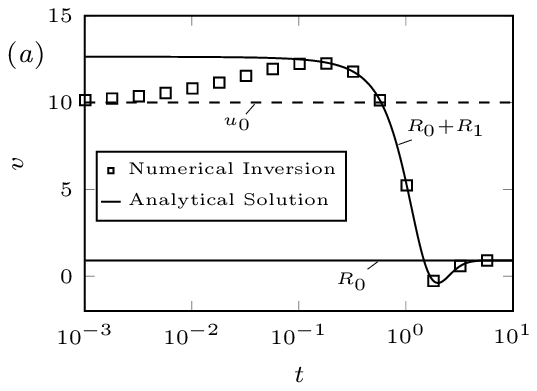}
                \label{fig:knt_v_valid}
                }
            \subfigure{
                \includegraphics{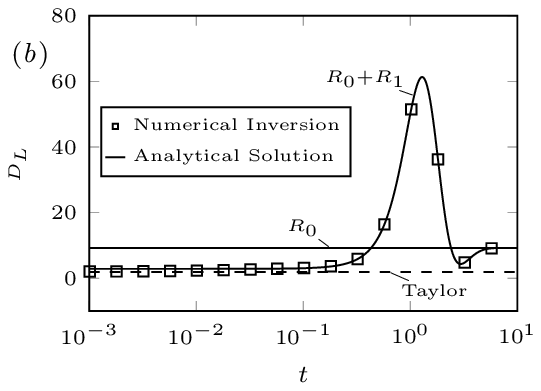}
                \label{fig:knt_D_valid}
            }
            \caption{Comparison of the zeroth-order approximation ($R_0$) and the first-order approximation ($R_0 + R_1$) with the numerical inversion of the Laplace transform (squares) using Talbot's method. \cor{$R_0$ and $R_1$ are the zeroth and first residues of the moments defined in equation \eqref{eq:mn_Rk} and \eqref{eq:Rk_def}}. Results are shown for $\Pe = 10, \ka = 10, \kd = 1$. Panel (\textit{a}) shows the transport velocity $v$, where the mean flow velocity $u_0=U_0 H/D=\Pe$. Panel (\textit{b}) shows the dispersion coefficient $D_L$, where the dashed line labelled as Taylor denotes the Taylor dispersion $2/105 \Pe^2$.}
            \label{fig:knt_valid}
        \end{figure}

         \cor{The zeroth- and first-order terms corresponds to the residues $R_0$ and $R_1$ in \eqref{eq:mn_Rk}.} Figure \ref{fig:knt_valid}(\textit{a}) and \ref{fig:knt_valid}(\textit{b}) show the comparison of the zeroth-order approximation $R_0$ and first-order approximation $R_0 + R_1$ with the numerical inversion of Laplace transform using Talbot's method \citep{Abate2006, Tucker2013}. As expected, the first-order approximation $R_0 + R_1$ captures the solution at both the early and the late times, while the zeroth-order approximation $R_0$ only describes the late time behavior. Additional tests show that the first-order approximation is sufficient to describe the solution for a large range of $\ka$ and $\kd$ \cor{after a short initial time}. Therefore, we truncate the series solution (\ref{eq:moment_series}) by retaining only the zeroth- and first-order terms,
            \begin{subequations} \label{eq:fst_app}
                \begin{align}
                    m_0 & = a_0 + a_1 \ept1, \label{eq:fst_app_m0} \\
                    m_1 & = \bigl( \bk01 + \bk02 t \bigr) \, + \, \bigr( \bk11 + \bk12 t \bigr) \ept1,\\
                    m_2 & = \bigl( \ck01 + \ck02 t + \ck03 t^2 \bigr) \, + \, \bigl( \ck11 + \ck12 t + \ck13 t^2 \bigr) \ept1,
                \end{align}
            \end{subequations}
        where the higher-order terms describing the very early time behavior are ignored.

\section{Regimes of transport} \label{sec:results}
    In this section, we discuss the transition from the early fast transport to the late slow transport. Numerical simulations of the full two-dimensional problem illustrate the physical mechanism that leads to this transition. The truncated analytical solution provides the estimates of the associated time scales. First, we will use the simpler equilibrium sorption model to discuss the regime transition, followed by the more general kinetic case.

    \subsection{Two-dimensional simulations}
\begin{figure}
            \centering
            \includegraphics{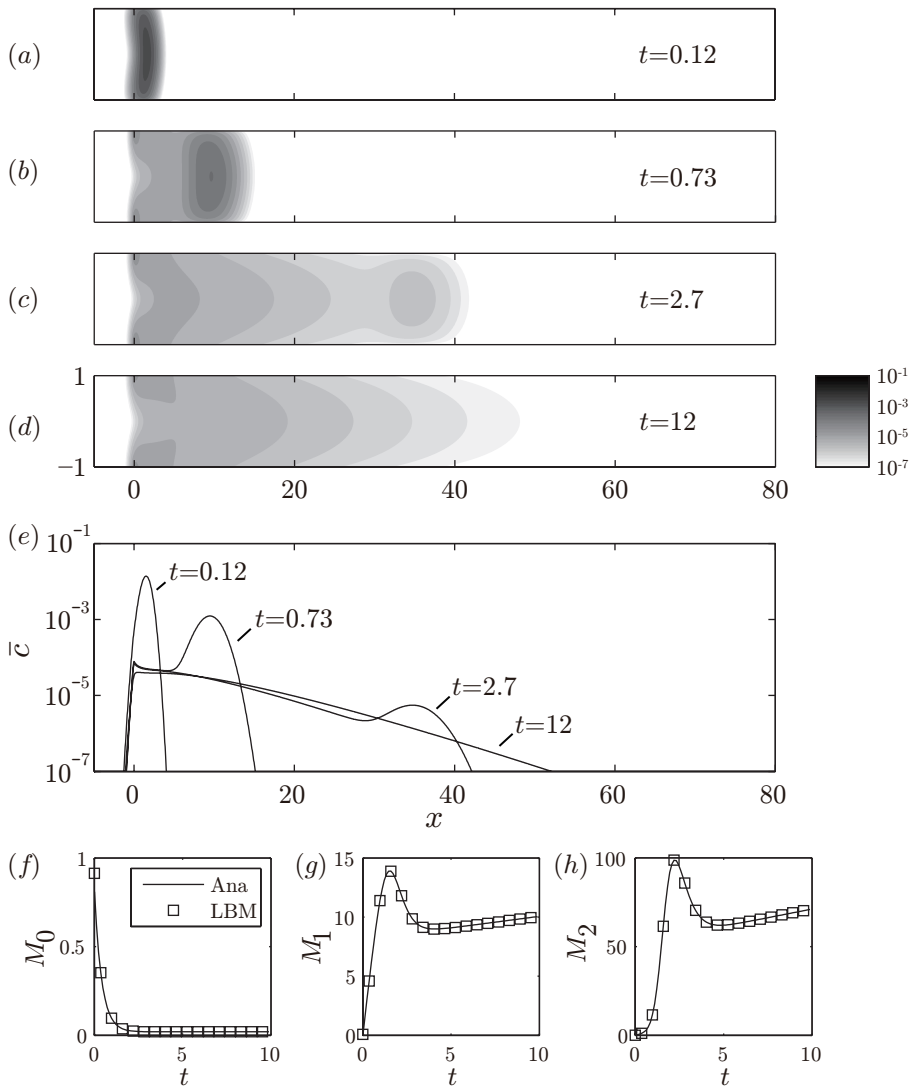}
            \caption{Two dimensional simulation of solute transport with sorption in Poiseuille flow with $\Pe = 10$, $\ka = 50$ and $\kd = 1$. Panel (\textit{a}--\textit{d}) show concentration distribution and panel (\textit{e}) shows transversely-averaged concentration profile at $t = 0.12, \, 0.73, \, 2.7, \,12$. \cor{Panel (\textit{f}--\textit{h}) show the evolution of the zeroth-, first- and second-order moments and compare the numerical simulation (LBM) with the first-order approximation of the analytical solution (Ana).}}
            \label{fig:2d_simulation}
        \end{figure}

        Figure \ref{fig:2d_simulation} shows two-dimensional simulations of the solute concentration at different times for $\Pe = 10$, $\ka = 50$ and $\kd = 1$. The full problem is numerically solved by the Lattice Boltzmann Method (LBM) \citep{Chen1998, Wang2010a, Zhang2015b}. The $\delta$-function initial condition is approximated by a piecewise constant function that is non-zero in a small interval around the origin. This approximation of the initial condition only affects the results in a short diffusive transient \cor{and the results agree well with the analytical solution (figure \ref{fig:2d_simulation}\textit{f}-\ref{fig:2d_simulation}\textit{h}).} 

        Initially, the strong adsorption removes the solute from the slow-moving fluid near the wall. The remaining solute in the center of the channel forms a fast-moving pulse (figure \ref{fig:2d_simulation}\textit{b} and \ref{fig:2d_simulation}\textit{c}), particularly evident in the transversely-averaged concentration shown in figure \ref{fig:2d_simulation}(\textit{e}). This corresponds to the increased solute transport velocity in the irreversible sorption case \citep{Lungu1982}. This regime persists as long as adsorption dominates.
        
        However, the fast-moving pulse decays rapidly and eventually desorption releases solute in its wake (figure \ref{fig:2d_simulation}\textit{d}). As the amount of desorbed solute in the slow-moving fluid near the wall increases, the solute transport velocity declines. This process continues until desorption at the back balances adsorption at the front. The transport velocity and dispersion coefficient will approach the slow transport described by the one-dimensional model of the transversely-averaged concentration in the reversible sorption case \citep{Khan1962}.

        
    \subsection{Equilibrium sorption model} \label{subsec:eq}
        \cor{Following the solution procedure in section \ref{sec:eq_sol},  this section presents results and analysis for equilibrium sorption model, $\gamma = \kp c$.} To demonstrate the different transport behaviors, we define a normalized transport velocity $\nmv$ and a normalized dispersion coefficient $\nmD$ as
        \begin{subeqnarray}
        \gdef\thesubequation{\theequation \textit{a,b}}
                \nmv = \frac{v}{Pe} \quad \mathrm{and} \quad \nmD = \frac{D_L - 1}{D_{t}}, \label{eq:nmD_def}
        \end{subeqnarray}
        where $D_t = 2/105 \Pe^2$ is the Taylor dispersion coefficient for a tracer in Poiseuille flow and \cor{the unit contribution of diffusion has been subtracted in the numerator of (\ref{eq:nmD_def}\textit{b}).}
        In this way, $\nmv > 1$ ($ \nmD > 1$) means increased velocity (dispersion) relative to a nonreactive tracer while $\nmv < 1$ ($\nmD < 1$) means decreased velocity (dispersion).

        

\begin{figure}
            \centering
            \subfigure{    
                \includegraphics{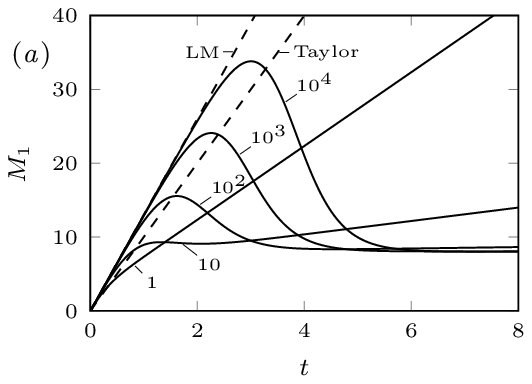}
                \label{fig:eq_M1}
                }
            \subfigure{ 
                \includegraphics{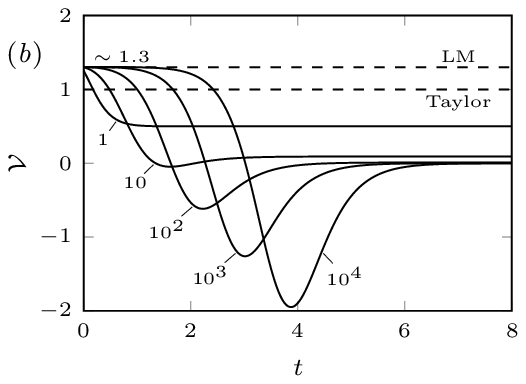}
                \label{fig:eq_v_series}
                }     
            \caption{Evolution of (\textit{a}) center of mass $M_1$ and (\textit{b}) normalized transport velocity $\nmv$ for different partition coefficients $\kp$. Dashed lines with labels LM and Taylor stand for the asymptotic regime of an adsorption-only case \citep{Lungu1982} and the asymptotic regime of a nonreactive tracer, respectively.}
            \label{fig:eq_M1_v}
        \end{figure}

        Figure \ref{fig:eq_M1_v} shows the evolution of the position of the center of mass $M_1$ and the normalized transport velocity $\nmv$ for different partition coefficients. For $\kp > 10$, a linear region emerges at early times in figure \ref{fig:eq_M1_v}(\textit{a}), corresponding to an initial plateau in figure \ref{fig:eq_M1_v}(\textit{b}). This corresponds to the well-developed early regime characterized by fast transport, approaching an asymptotic velocity $ 1 + 3/\upi^2 \approx 1.3$. This is consistent with the results in an adsorption-only case with $\ka \rightarrow \infty$ \citep{Lungu1982}. After a transition period, a second linear region at late times appears, corresponding to the decreased transport velocity $1/(1 + \kp)$.

\begin{figure}
            \centering
            \subfigure{
                \includegraphics{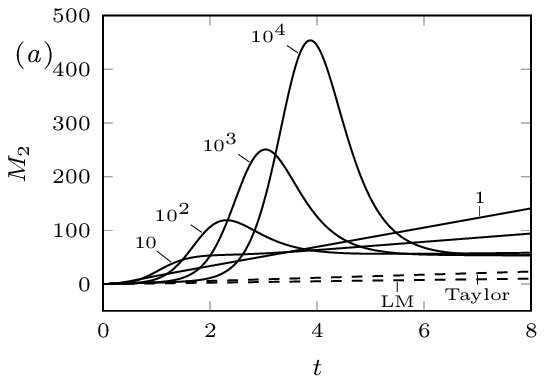}
                \label{fig:eq_M2}
                }
            \subfigure{    
                \includegraphics{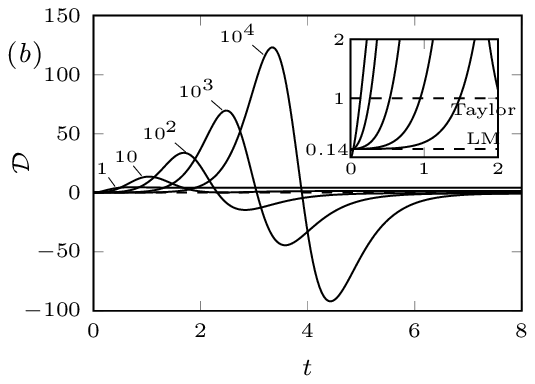}
                \label{fig:eq_D_series}
                }        
            \caption{Evolution of (\textit{a}) variance of solute mass distribution $M_2$ and (\textit{b}) the normalized dispersion coefficient $\nmD$ for different partition coefficients $\kp$. Dashed lines with labels LM and Taylor stand for the asymptotic regime of an adsorption-only case \citep{Lungu1982} and the asymptotic regime of a nonreactive tracer, respectively.}
            \label{fig:eq_M2_D}
        \end{figure}

        Figure \ref{fig:eq_M2_D} shows similar behaviors of the variance of solute mass in the fluid $M_2$ and the normalized dispersion coefficient $\nmD$. In the early regime, the dispersion coefficient is reduced relative to a tracer with $\nmD \sim 0.14$, which also agrees with the adsorption-only case with $\ka \rightarrow \infty$. In the late regime, the dispersion coefficient is given by the first two terms of (\ref{eq:late_D}), first obtained by \citet{Golay1958}. 

        Between the early and late time, there is a drastic transition of the transport behavior. Especially when $\kp$ is large, both $M_1$ and $M_2$ decrease after reaching maxima during the transition and this leads to negative velocity and dispersion coefficient. Physically, it means that desorption near the origin dominates over the fast-moving pulse in figure \ref{fig:2d_simulation} so that the center of mass shifts backwards and the variance reduces because the transversely-averaged concentration distribution changes from a bimodal type (one peak near the origin and the other at the pulse front) to a unimodal type (single peak near the origin).

        To compare the early and late behaviors as a function of $\kp$, we define the normalized \del{asymptotic} early and late velocities as
        \begin{subeqnarray}
        \gdef\thesubequation{\theequation \textit{a,b}}    
            \nmv_e = \frac{v_1}{\Pe} \quad \mathrm{and} \quad \nmv_l = \frac{v_0}{\Pe},
        \end{subeqnarray}
        and similarly we define normalized early and late dispersion coefficients as
        \addtocounter{equation}{-1}
        \begin{subeqnarray}
        \gdef\thesubequation{\theequation \textit{c,d}}
                \nmD_e  = \frac{D_1-1}{D_t} \quad \mathrm{and} \quad \nmD_l  = \frac{D_0 - 1/(1+\kp)}{D_t},
        \end{subeqnarray}
        where $v_0$, $D_0$ and $v_1$, $D_1$ are obtained from the zeroth- and first-order terms of the solution, the equilibrium limits of (\ref{eq:late_v_D}) and (\ref{eq:early_v_D}). Note that at early times the effective diffusion is not affected by sorption while it is reduced by a factor of $1/(1+k)$ at late times.
        
\begin{figure}
                \centering
                \subfigure{    
                    \includegraphics{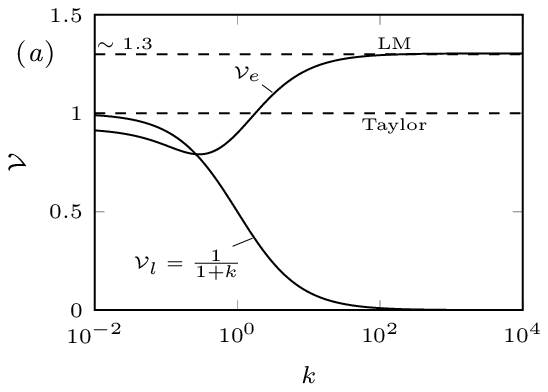}
                    \label{fig:eq_early_late_v}
                    }
                \subfigure{    
                    \includegraphics{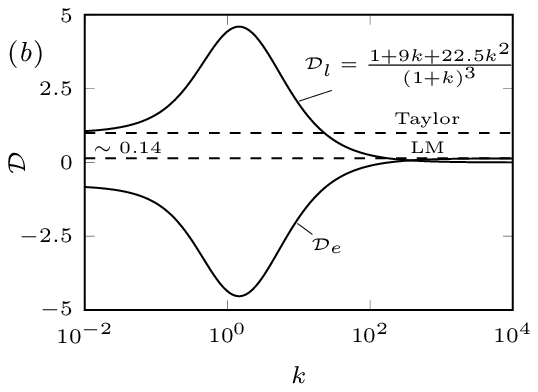}
                    \label{fig:eq_early_late_D}
                    }        
                \caption{(\textit{a}) Normalized \del{asymptotic} early velocity $\nmv_e = v_1/\Pe$, late velocity $\nmv_l = v_0/\Pe$ and (\textit{b}) normalized early dispersion coefficient $\nmD_e = (D_1-1)/D_t $, late dispersion coefficient $\nmD_l = (D_0 - 1/(1+\kp))/D_t$ as a function of partition coefficient $\kp$.}
                \label{fig:eq_early_late}
            \end{figure}

        As shown in figure \ref{fig:eq_early_late}, for large $\kp$ the difference of normalized transport velocity between $\nmv_e$ and $\nmv_l$ is the largest and the normalized early-time dispersion coefficient $\nmD_e$ asymptotes to $0.14$. The normalized late-time dispersion coefficient $\nmD_l$ first increases with $\kp$, then reduces towards $0$. \cor{For small $\kp$, the early velocity $\nmv_e$ and dispersion coefficient $\nmD_e$ don't reach the asymptotic values $1.3$ and $0.14$. In this case, the early regime is not well-developed and the first-order terms of the solution are not dominant.} Therefore, $\nmv_e$ and $\nmD_e$ don't represent the transport behavior in this case.

        For a tracer in Poiseuille flow, the preasymptotic transport before equilibrium has been studied extensively \citep[e.g.][]{Gill1970, Haber1988, Mercer1990, Latini2001, Dentz2007, Bolster2011, Wang2012}. Typically, diffusion dominates when $t \ll t_d = \Pe^{-2/3}$ and after the characteristic equilibrium time scale, $t = 1$, solute transport can be described by the transversely-averaged model with the mean flow velocity and the dispersion coefficient $D_t$. However, for a reactive case considered here, the time scale to reach equilibrium can be quite different from the tracer case because surface reactions introduce additional characteristic time scales.

        In the first-order approximation (\ref{eq:fst_app}), a series of time scales can be defined by comparing the zeroth-order and first-order terms. Take $m_0$ as an example, by comparing $a_0$ and $a_1 \ept1$, we can define a time scale  as
            \begin{equation} \label{eq:early_time_ana}
                \cor{t_1 = \frac{1}{p_1^2} \ln \frac{a_1}{a_0}  =  \frac{1}{p_1^2} \ln \frac{2 \kp^2 (\kp + 1)}{\kp^2 p_1^2 + \kp + 1} .}
            \end{equation}
        This time scale indicates the transition from the early regime when the transport is dominated by the first-order terms to the late regime dominated by the zeroth-order terms.
        Similarly, other time scales can be determined by comparing $b_k$ and $c_k$. We notice that the coefficients in the series solution have the following property
            \begin{equation}
                \frac{a_1}{a_0} < \frac{\bk12}{\bk02} \sim \frac{\ck12}{\ck02}< \frac{\ck13}{\ck03},
            \end{equation}
        which means that $t_1$ will give the smallest time scale. At the same time, since the time scales for the velocity determined by $\bk12/\bk02$ and the dispersion coefficient determined by $\ck12/\ck02$ have the same scaling, we can choose
            \begin{equation} \label{eq:late_time_ana}
                \cor{t_2 = \frac{1}{p_1^2} \ln \frac{\bk12}{\bk02} = \frac{1}{p_1^2} \ln \frac{\kp^2 (\kp+1)^2 (4\kp^2 p_1^4 + 3\kp^2p_1^2 - 3\kp + 4p_1^2 - 3)}{2 p_1^2 (\kp^2 p_1^2 + \kp + 1)^2}}
            \end{equation}
        as a critical time scale, after which the zeroth-order terms dominate and the late time behavior emerges. \cor{Note that both of the time scales are not dependent on $\Pe$.} Consequently, the transient solute transport with sorption can be divided into the following three regimes,
        
            \begin{tabular}{lll}
                (I) & \del{$t_d < t < t_1$} \cor{$0 < t < t_1$} & : early regime with fast transport  \\
                (II) & $t_1 < t < t_2$ & : transition period \\
                (III) & $t_2 < t$  & : late regime with slow transport
            \end{tabular}
        
\begin{figure}
              \centerline{\includegraphics{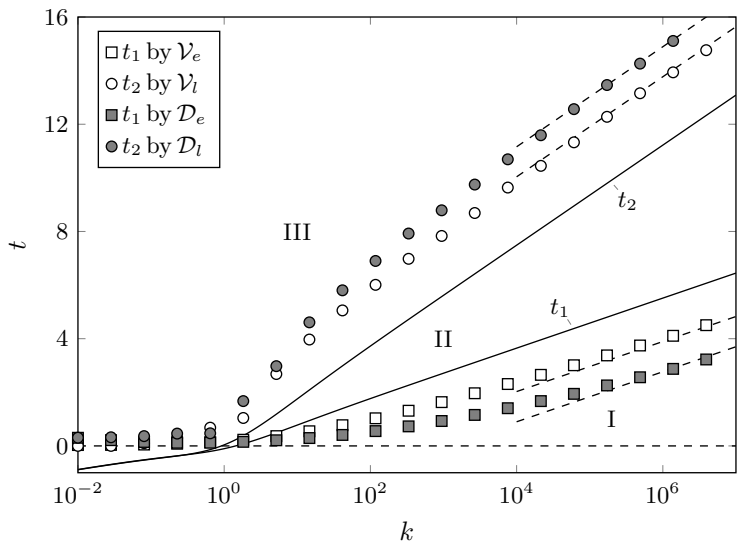}}
              \caption{Variation of the early time scale $t_1$ and the late time scale $t_2$ as functions of different partition coefficient $\kp$. \cor{The solid lines show $t_1$ and $t_2$ from (\ref{eq:early_time_ana}) and (\ref{eq:late_time_ana}), respectively and the dashed lines indicate the scalings for large $\kp$ given by \eqref{eq:time_scale_limit}, which have been shifted to match the symbols determined by the numerical inverse Laplace transform. White symbols are determined by velocity and gray symbols are determined by dispersion coefficient}. The transient solute transport with sorption \del{after the initial diffusive regime, $t > t_d$,} is divided into three regimes: (I) early regime with fast transport; (II) transition period; (III) late regime with slow transport.
              }
            \label{fig:eq_time}
            \end{figure}

        As shown in figure \ref{fig:eq_M1_v} and \ref{fig:eq_M2_D}, the duration of the early regime, as well as the transition period, increases with increasing $\kp$. In the limit of large $\kp$, we have
            \begin{subeqnarray}
            \gdef\thesubequation{\theequation \textit{a,b}}
            \label{eq:time_scale_limit}
                    \lim_{\kp \rightarrow \infty} t_1 = \frac{4}{\upi^2} \ln \kp \quad \mathrm{and} \quad
                    \lim_{\kp \rightarrow \infty} t_2 = \frac{8}{\upi^2} \ln \kp,
            \end{subeqnarray}
        where we have used $\lim_{\kp \rightarrow \infty} p_1 = \upi/2$. Equations (\ref{eq:time_scale_limit}) predict a linear relationship between $t_1$, $t_2$ and $\ln{\kp}$ when $\kp$ is large and $t_2 \sim 2\, t_1$. Figure \ref{fig:eq_time} compares results from the numerical inverse Laplace transform with these analytically determined time scales as a function of $\kp$. When $\kp \gg 1$, $t$ scales with $\ln \kp $, as predicted by (\ref{eq:time_scale_limit}). \cor{For small $\kp$, the early regime is so short that it is generally not observed.} 

    \subsection{Kinetic sorption model}
        In the kinetic sorption model, there are two additional governing parameters, namely, the dimensionless adsorption rate constant $\ka$ and the dimensionless desorption rate constant $\kd$. Generally, a similar transition from early to late behavior can be observed and the equilibrium results are recovered when kinetics are fast, i.e. $\ka \gg 1$ and $\kd \gg 1$.
        
\begin{figure}
                \centering
                \subfigure{
                    \includegraphics{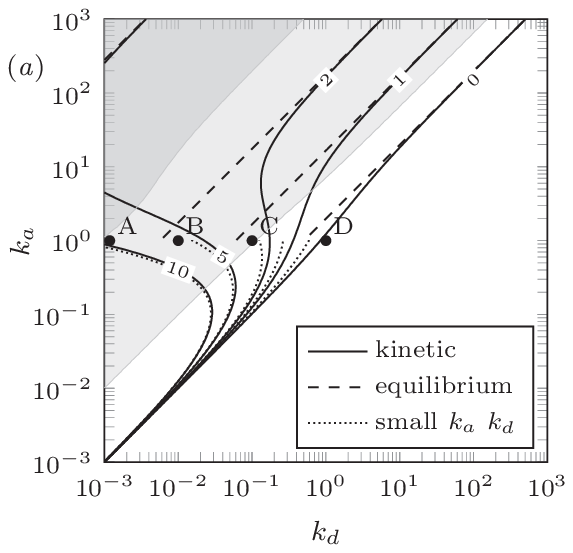}
                    \label{fig:knt_early_t}
                    }
                \subfigure{    
                    \includegraphics{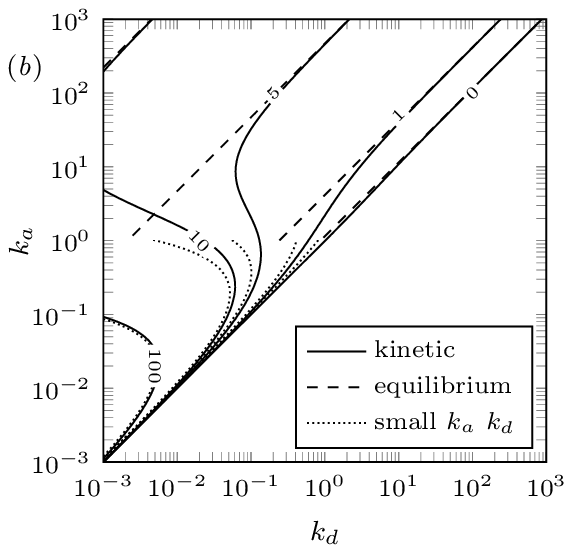}
                    \label{fig:knt_late_t}
                    }
                 \subfigure{                    \includegraphics{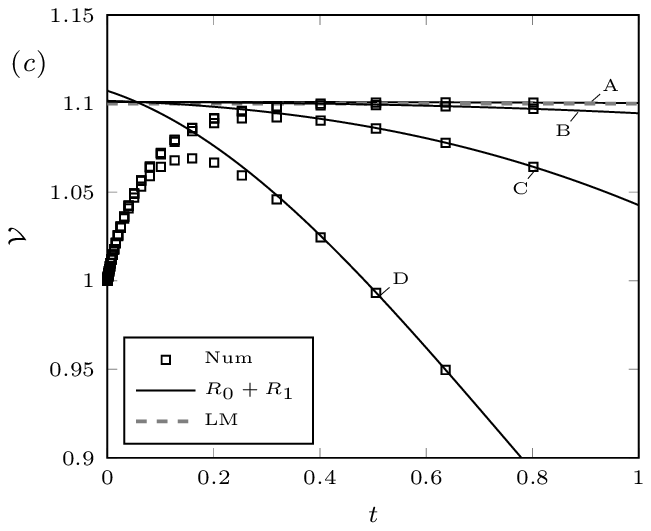}
                     \label{fig:knt_v_well_develp}
                    }
                \caption{Contour lines of (\textit{a}) the early time scale $t_1$ and (\textit{b}) the late time scale $t_2$ for the kinetic sorption model. Dashed lines represent the time scales obtained from the equilibrium sorption model for large $\ka$ and $\kd$ and dotted lines represent approximations for small $\ka$ and $\kd$, given by (\ref{eq:small_kakd_app}).  \cor{Shaded area in (\textit{a}) represents the region where the early regime is well-developed, i.e., the velocity (light-shaded area) and dispersion coefficient (dark-shaded area) are close to the limiting values given by the adsorption-only case.} Panel(c) shows the evolution of velocity at early time for the conditions labeled as A, B, C, D in panel(a). Symbols  are  the  results  from  full  numerical inverse Laplace transform, solid lines are first-order approximation of the analytical solution and dashed line is the asymptotic value for $\kd = 0$ at $\ka = 1$, given by \citet{Lungu1982}.}
                \label{fig:knt_diagram}
            \end{figure}

        Similar to the way that the time scales are determined for the equilibrium model, we can obtain $t_1$ and $t_2$ for the kinetic model using \eqref{eq:early_time_ana} and \eqref{eq:late_time_ana}, 
		    \cor{\begin{subequations}
		        \begin{align}
		            t_1 & = \frac{1}{p_1^2} \ln \frac{2 \ka^2 (\ka + \kd)}{\kd \bigl(p_1^4 + (\ka^2+\ka -2\kd) p_1^2 + \ka \kd + \kd^2 \bigr)}, \\
		            t_2 & = \frac{1}{p_1^2} \ln \frac{\ka^2 (\ka + \kd)^2 \bigl( 4 p_1^6 + (4\ka^2- 8\kd -3) p_1^4 + (3 \ka^2 + 3\ka +4\kd^2 +6\kd) p_1^2 
		             - 3\ka \kd   - 3\kd^2 \bigr) }{2\kd^2 p_1^2 \bigl( p_1^4 + (\ka^2 + \ka - 2\kd) p_1^2 + \ka \kd + \kd^2 \bigr)^2},
		        \end{align}
		    \end{subequations}}
        which are shown in figure \ref{fig:knt_diagram}.
        
        The early regime is only observed when $\ka$ exceeds $\kd$. In all other cases, \del{the diffusive regime is immediately followed by the late regime} \cor{transition occurs from the very beginning followed by a dominated late regime}. When both $\ka$ and $\kd$ are large, the time scales of the kinetic model recover those of the equilibrium model. 
        
        However, if the rates decrease, the kinetic time scales become longer. In this case, the root $p_1$ of (\ref{eq:tr_p_knt}) can be approximated by $p_1^2 \approx \ka + \kd$ using Taylor expansion for $\tan(p)$. Then the ratios  $a_1/a_0$ and $\bk12/\bk02$ used to obtain the time scales simplify to
            \begin{subeqnarray} \label{eq:small_kakd_app}
            \gdef\thesubequation{\theequation \textit{a,b}}
                    \frac{a_1}{a_0} = \frac{2 \ka}{\kd (\ka + 2)} \quad \mathrm{and} \quad
                    \frac{\bk12}{\bk02}  = \frac{\ka^2 (4\ka + 4\kd + 7)}{2 \kd^2 (\ka + 2)^2}.
            \end{subeqnarray}
        This analysis shows that both time scales increase dramatically in the lower left region \cor{where $\ka$ and $\kd$ are small} in figure \ref{fig:knt_diagram}. In this region, the duration of the early regime is long, but the deviations of velocity and dispersion coefficient from the tracer case are minor \cor{as the limiting values given by the adsorption-only case approach unity with small $\ka$}.
        Physically, this region corresponds to a kinetically slow-sorbing ($\ka, \kd \ll 1$) solute with a large partition coefficient $\kp \gg 1$.        
        

        The analytical solution presented in \textsection \ref{sec:degeneration} recovers the previous analysis in the limit of $\kd = 0$ \citep{Lungu1982}. \cor{This limiting solution puts an upper bound on the transport velocity and a lower bound on the dispersion coefficient in the early regime. If the early regime is well-developed, the limiting solution given by \citet{Lungu1982} provides a good approximation for finite $\kd$, see figure \ref{fig:knt_v_well_develp}. The well-developed early regime is indicated by gray shadings in figure \ref{fig:knt_diagram}(\textit{a}), where the early-time asymptotic transport velocity and dispersion coefficient given by \eqref{eq:early_v_D} are within 10\% of the limiting values given by \citet{Lungu1982}. 
        Case A, B, C give examples of well-developed early regime, while the velocity doesn't reach the asymptotic value in case D.
        Generally, if $\kp > 10$ ($> 1000$), the early-time velocity (dispersion coefficient) are well developed. Note that the first-order analytical solution for transport velocity ($R_0 + R_1$) shown in figure \ref{fig:knt_diagram}(\textit{c}) is computed by \eqref{eq:v_D_def} and \eqref{eq:fst_app}.}
        


\section{Conclusion}

    In this work, we reconcile two different analyses of solute transport with sorption in Poiseuille flow that reached apparently contradictory conclusions. We show that these two analyses capture different regimes of the transport. Generally, the solute experiences an early regime with fast transport velocity if adsorption dominates desorption. At late times, when desorption becomes important, the solute transport slows down. This leads to a regime transition that scales as $\ln \kp$ for the equilibrium sorption model, where $\kp$ is the dimensionless partition coefficient. Therefore, the early regime is more pronounced when $\kp$ is large. In the kinetic sorption model, the early regime is also observed if the kinetics are slow and the dimensionless adsorption rate constant $\ka$ exceeds the dimensionless desorption rate constant $\kd$. \cor{As long as $\ka \gg \kd$, the early regime is well-developed and the transport velocity and the dispersion coefficient in this early regime are well approximated by the analysis of \citet{Lungu1982} in the limit of $\kd = 0$.}
    
    The time scales presented in this work allow the determination of the dominant transport behavior for a given application. Experience shows that the late regime dominates the subsurface transport of sorbing contaminants \cor{in fractures}. However, the early regime may be important in the biomedical applications where transport occurs over smaller distances. Our analysis may also allow a design of chromatography columns that can achieve opposite separation results.

    

\begin{acknowledgments}
    L. Zhang and M. Hesse are grateful to Prof. Howard Stone and Dr. Zhong Zheng for helpful discussions, which motivate this work. The authors are also grateful to Prof. Howard Stone for carefully reading the manuscript. L. Zhang acknowledges the financial support by China Scholarship Council's (CSC) Chinese Government Graduate Student Oversea Study Program. M. Wang acknowledges the financial support by the NSF grant of China (No.51676107, 91634107, U1562217), National Science and Technology Major Project on Oil and Gas (No.2017ZX05013001).
\end{acknowledgments}

\appendix 
     \section{Effect of initial condition} \label{subsec:init_cd} 
        The initial distribution of solute mass manifest itself either as a source term, $c_n^*(t=0)$, or a constant in the boundary condition, $\gamma_n^*(t=0)$, in the ODE system (\ref{eq:ode}). These effects can be important in our problem in the sense that it may affect the form of the solutions of the moments. General discussion on this topic is out of the scope of this paper, and we show a special case as an example.
        
         In the previous formulation, we assume initially there is no mass adsorbed on the wall, $\gamma_n^*(t=0) = 0$. In this section, we change the initial condition by retaining the uniform release in the fluid, but assuming the mass distribution between the wall and the bulk has reached equilibrium, namely, $c(t=0) = \delta(x)/(1+\kp)$, $\gamma(t=0) = \delta(x) \kp/(1+\kp) $. Following the same procedure in \textsection \ref{subsec:me_lt} and \textsection \ref{subsec:ilt}, we find that the moments in Laplace space $\skew3\hat{m}_0, \skew3\hat{m}_1, \skew3\hat{m}_2$ are no longer in the form of (\ref{eq:moment_gnr}), but with a slight difference, 
             \begin{subequations} \label{eq:moment_gnr_ieq}
                \begin{align}
                    \skew3\hat{m}_0(s) & = \frac{Q_0 (s) }{s}, \\
                    \skew3\hat{m}_1(s) & = \frac{Q_1 (s) }{s \,\tr(s)}, \\
                    \skew3\hat{m}_2(s) & = \frac{Q_2 (s) }{s \,\tr^{2} (s)},                    
                \end{align}
            \end{subequations}
        where $Q_0$, $Q_1$ and $Q_2$ are different from $N_0$, $N_1$ and $N_2$. In fact, $Q_0 = 1/(1+k)$. Essentially, the order of the all the singularities, other than the zeroth-order, reduce one in the solutions. Therefore, the series solutions obtained by residue theorem are written as
            \begin{subequations} \label{eq:moment_series_ieq}
                \begin{align} 
                    m_0(t) & = a_0, \\
                    m_1(t) & = \bk01 + \bk02 t + \sum\limits_{k=1}^{\infty} \skew3\tilde{b}_k^{(1)} \exp(-p_k^2 t) \\
                    m_2(t) & = \ck01 + \ck02 t + \ck03 t^2 + \sum\limits_{k=0}^{\infty} \skew3\tilde{c}_k^{(1)} \exp(-p_k^2 t) + \skew3\tilde{c}_k^{(2)} t \exp(-p_k^2 t).
                \end{align}
            \end{subequations}        
        where $\skew3\tilde{b}_k^{(1)}$, $\skew3\tilde{c}_k^{(1)}$ and $\skew3\tilde{c}_k^{(2)}$ are different from $b_k^{(1)}$, $c_k^{(1)}$ and $c_k^{(2)}$. Note that the long-time velocity and dispersion coefficient determined by $a_0$, $\bk02$, $\ck02$ don't change. However, since $b_k^{(2)}$ diminishes, the early regime will not be well-developed in this case. 
        Physically, the solute that is initially adsorbed onto the wall begins to desorb much earlier, and hence reduces the duration of the early regime. In the limit of $\kd=0$, $k \rightarrow \infty$ and the initial solute mass in the fluid $c(t=0)$ vanishes so that the results by \citet{Lungu1982} can not be properly recovered in this case.

\bibliographystyle{jfm}
\bibliography{Reference}

\end{document}